\documentclass[%
aip,
reprint,
superscriptaddress,
ymb,
prb,
twocolumn,
titlepage,
floatfix,
showpacs,
]{revtex4-2}

\usepackage{graphicx}
\usepackage{hyperref}
\usepackage{amssymb}
\usepackage{amsmath}
\usepackage{textcomp}
\usepackage{xcolor}

\allowdisplaybreaks

\definecolor{linkblue}{RGB}{49,49,148}
\hypersetup{linkcolor  = linkblue,citecolor  = linkblue,urlcolor   = linkblue,colorlinks = true,}

\makeatletter
\renewcommand*{\eqref}[1]{%
  \hyperref[{#1}]{\textup{\tagform@{\ref*{#1}}}}%
}
\makeatother

\begin{document}
\preprint{AIP/123-QED}

\title{
Coherent control of magnetization precession by double-pulse activation of effective fields from magnetoacoustics and demagnetization}

\author{M.~Mattern}
\affiliation{Institut f\"ur Physik \& Astronomie,  Universit\"at Potsdam, 14476 Potsdam, Germany}

\author{F.-C.~Weber}
\affiliation{Institut f\"ur Physik \& Astronomie,  Universit\"at Potsdam,  14476 Potsdam, Germany}

\author{D.~Engel}
\affiliation{Max-Born-Institut für Nichtlineare Optik und Kurzzeitspektroskopie, 12489 Berlin, Germany}

\author{C.~Korff-Schmising}
\affiliation{Max-Born-Institut für Nichtlineare Optik und Kurzzeitspektroskopie, 12489 Berlin, Germany}

\author{M.~Bargheer}
\email{bargheer@uni-potsdam.de}
\affiliation{Institut  f\"ur Physik \& Astronomie, Universit\"at Potsdam, 14476 Potsdam, Germany}
\affiliation{Helmholtz-Zentrum Berlin f\"ur Materialien und Energie, 12489 Berlin, Germany}

\date{\today}

\begin{abstract}
We demonstrate the coherent optical control of magnetization precession in a thin Ni film by a second excitation pulse which amplifies or attenuates the precession induced by a first pulse depending on the fluences of the pump-pulses and the pump-pump delay. This control goes beyond the conventional strategy, where the same mechanism drives the precession in or out-of phase. We balance the magneto-acoustic mechanism driven by quasi-static strain and the shape-anisotropy change triggered by laser-induced demagnetization. These mechanisms tilt the transient effective magnetic field in opposite directions in case of negative magneto-elastic coupling ($b_1<0$). While the strain response is linear in the fluence, demagnetization is nonlinear near the Curie temperature, enabling fluence-based control scenarios.
\end{abstract}

\maketitle

Since the discovery of ultrafast demagnetization in thin Ni films\cite{beau1996}, the field of ultrafast magnetization dynamics has evolved from observing ultrafast magnetization dynamics \cite{stam2014,sand2017} to controlling the magnetization on ultrafast timescales\cite{kova2013, razd2017, vedm2020, stup2021, igar2023}.

The fastest ways to coherently change the magnetization state is the excitation of precessional motion reported for a wide range of ferro- or ferri-magnetic metals\cite{kamp2002a, bigo2005, ma2015}, semiconductors\cite{sche2010} and insulators \cite{deb2016, deb2018}. Depending on the magnetically ordered material, the precession can be driven by temperature-induced changes of the magnetocrystalline anisotropy \cite{ma2015, bigo2005}, ultrafast demagnetization affecting the shape anisotropy \cite{ma2015, kamp2002a, shin2023, jare2023} or by both local quasi-static expansion \cite{shin2022, shin2023} and propagating strain pulses \cite{sche2010, bomb2013, kim2015, kim2017, jare2023} via material-specific magneto-elastic coupling mediated by spin-orbit coupling.

In classical coherent control experiments \cite{zhan2002,kamp2011,kim2015,yang2018,deb2022} the magnetization precession is tuned by the pump-pump delay in a double-pulse excitation scheme, where both excitations induce the same driving mechanism, which depends linearly on the fluence. Very recently, Shin and co-workers\cite{shin2022, shin2023} controlled the precession amplitude and phase via the composition of a Ni$_x$Fe$_{100-x}$ film, which tunes the magneto-elastic coupling $b_1$, the Curie-temperature $T_\text{C}$ and the saturation magnetization $M_\text{sat}$. This changes the dominant driving mechanism from quasi-static expansion to laser-induced demagnetization that counteract in materials with negative magneto-elastic coupling ($b_1<0$). In addition, they showed this counterbalance to depend on the fluence due to the non-linear increase of the demagnetization near $T_\text{C}$ that becomes the dominant driving mechanism and switches the precession phase by $\pi$.

In this letter, we combine coherent control of the magnetization precession by double-pulse excitation with tuning of the dominant driving mechanism via the fluences of both pump-pulses and the pump-pump delay. The counterbalance of laser-induced demagnetization and the negative magneto-elastic coupling ($b_1<0$) in our $20\,\text{nm}$-thin Ni film strongly depends on the laser-induced temperature rise and the initial sample temperature since the demagnetization increases non-linearly near $T_\text{C}$. We demonstrate a transition of the dominant driving mechanism of the second excitation from magneto-elastic coupling to demagnetization in three scenarios: The temperature before the second pulse is tuned via the fluence of the first excitation or the pump-pump delay or, alternatively, the fluence of the second pulse determines the temperature rise. The change of the dominant driving mechanism switches the effect of the second pulse from amplifying to attenuating the precession induced by the first pulse, which opens rich opportunities to control the precession by double-pulse excitation.

\begin{figure}[t!]
\centering
\includegraphics[width =\columnwidth]{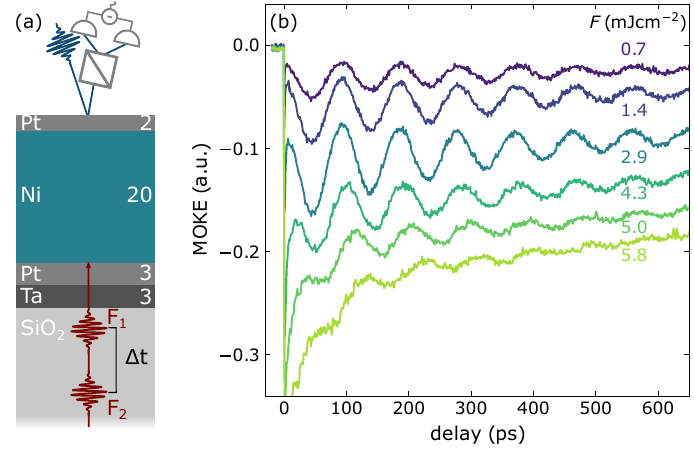}
\caption{\label{fig:fig_1_overview}\textbf{Fluence-dependent precession amplitude:} (a) Sketch of the sample structure containing a $20\,\text{nm}$-thick Ni layer and the time-resolved polar MOKE experiment providing double-pulse excitation with variable pump-pump delay $\Delta t$ and fluence of the first ($F_1$) and the second ($F_2$) pump pulse. (b) Fluence-dependent magnetization precession upon a single excitation for an external field of $350\,\text{mT}$ at $35^\circ$ with respect to the sample surface normal.}
\end{figure}

In our all-optical transient polar magneto-optical Kerr experiments (trMOKE) sketched in Fig.~\ref{fig:fig_1_overview}(a), a $20\,\text{nm}$ Ni film embedded within a $2\,\text{nm}$ Pt capping and $3\,\text{nm}$ Pt and Ta buffer layers is excited by two p-polarized $120\,\text{fs}$-long $800\,\text{nm}$ pump-pulses from the substrate-side at normal incidence with tunable pump-pump delay $\Delta t$ and individually adjustable fluences $F_1$ and $F_2$. We probe the transient out-of-plane magnetization of the Ni film by detecting the polarization rotation of a $400\,\text{nm}$ probe pulse that is reflected at near normal incidence. The external magnetic field of around $350\,\text{mT}$ is applied along $35^\circ$ with respect to the surface normal set by a rotatable permanent magnet\cite{jare2023} and is below the out-of-plane saturation field $\mu_0 H_\text{sat}=500\,\text{mT}$. 

Figure~\ref{fig:fig_1_overview}(b) displays the fluence-dependent transient out-of-plane magnetization upon a single excitation that induces demagnetization and drives a coherent magnetization precession. With increasing fluence, the out-of-plane magnetization is reduced more and more and we observe a critical slowing down of the remagnetization\cite{you2018}. While the amplitude of the demagnetization increases with increasing fluence, the amplitude of the coherent magnetization precession increases only for small fluences ($<2.9\,\text{mJ}\,\text{cm}^{-2}$). For further increased fluence the precession is diminished again, such that the amplitude for $5.0$ and $0.7\,\text{mJ}\,\text{cm}^{-2}$ become comparable.

This non-linear fluence dependence originates from the opposite effect that laser-induced expansion and demagnetization have on the orientation of the effective magnetic field in materials with negative magneto-elastic coupling \cite{shin2022}. The effective field $\mu_0 \Vec{H}_\text{eff}=-\nabla_M F\textsubscript{M}$ is derived from the Free energy $F_\text{M}$ of the macroscopic magnetization and determines the orientation of the magnetization vector $\vec{M}=M_\text{sat} \Vec{m}$. For our thin polycrystalline Ni film grown by magnetron sputtering on a PGO substrate this Free energy is given by a Zeeman energy originating from the external magnetic field $\Vec{H}_\text{ext}$, a shape anisotropy determined by the thin film geometry and a magneto-elastic contribution induced by the mean out-of-plane strain $\eta_\text{Ni}(t)$\cite{jare2023, shin2023}:
\begin{equation}
    \begin{split}
        F\textsubscript{M}(\Vec{m}, t) &= - \mu_0 M_\text{sat}\Vec{m}(t) \cdot \Vec{H}\textsubscript{ext} + \frac{\mu_0 M_\text{sat}^2}{2} m_z(t)^2 \\\
        &\textcolor{white}{=} \, + b_1\eta_\text{Ni}(t) \, m_z(t)^2 \;.
    \end{split}
\label{eq:eq_1_free_energy}
\end{equation}
The latter two contributions lead to effective fields that are proportional to the out-of-plane component of the magnetization vector $m_z$. According to Eq.~\eqref{eq:eq_1_free_energy}, ultrafast demagnetization, i.e. a rapid drop of $m_z(t)$, and the laser-induced strain response of Ni $\eta_\text{Ni}(t)=\eta_\text{qs}(t)+\eta_\text{sp}(t)$ composed of a quasi-static expansion $\eta_\text{qs}$ and propagating strain pulses $\eta_\text{sp}$ are the driving mechanisms of the magnetization precession. Since the round-trip time of the strain pulse through the metallic heterostructure of $\approx5\,\text{ps}$ is an order of magnitude shorter than the precession period, the contribution of $\eta_\text{sp}$ becomes negligible\cite{kim2017} and the laser-induced demagnetization and quasi-static expansion of Ni remain as driving mechanisms. Due to the negligible in-plane saturation field of $\mu_0 H_\text{sat}=5\,\text{mT}$ representing the magnetic easy axis, we also neglect any contribution from a magneto-crystalline anisotropy\cite{jare2023, shin2023}.

\begin{figure}[t!]
\centering
\includegraphics[width =\columnwidth]{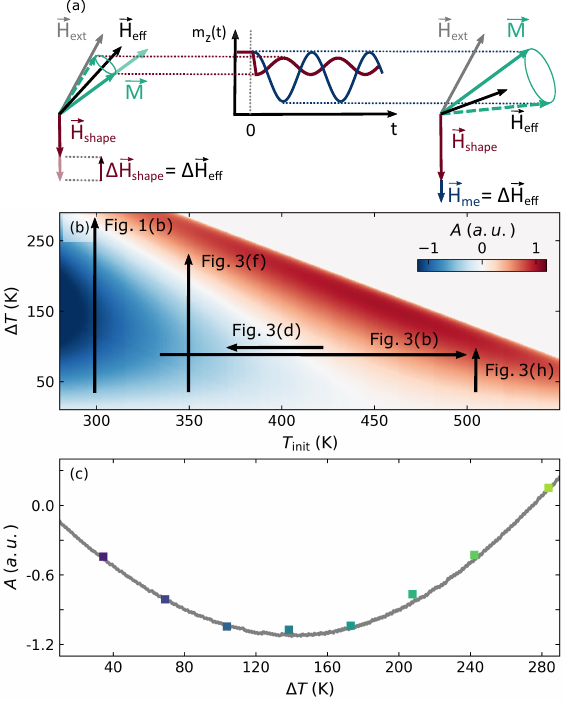}
\caption{\label{fig:fig_2_sim}\textbf{Modelled precession amplitude:} (a) Sketch of the counteracting effect of laser-induced demagnetization (left) and quasi-static expansion (right) on the effective field driving precession with a $\pi$ phase shift displayed by the red and blue solid line in the middle, respectively. (b) Amplitude of the precession $A$ as function of the laser-induced temperature rise in the phonons $\Delta T$ and the sample temperature before excitation $T_\text{init}$ modelled according to Eq.~\eqref{eq:eq_1_free_energy}. The arrows visualize the measurement series in Figs.~\ref{fig:fig_1_overview} and~\ref{fig:fig_3_data}. (c) The modelled precession amplitude as function of $\Delta T$ at $T_\text{init}=300\,\text{K}$ (solid line) corresponds to the left arrow in (b) and captures the measured amplitude of the fluence series in Fig.~\ref{fig:fig_1_overview}(b) (coloured symbols).}
\end{figure}

These mechanisms induce a change of the effective field $\Delta \Vec{H}_\text{eff}$ that adds to the effective field before excitation $\Vec{H}_{\text{eff},0}$, that determines the initial direction of the magnetization. This tilting of the effective field triggers precessional motion of $\Vec{M}$ according to the Landau-Lifshitz-Gilbert equation\cite{bigo2005}. Figure~\ref{fig:fig_2_sim}(a) sketches the resulting effective field upon laser excitation $\Vec{H}_\text{eff}=\Vec{H}_{\text{eff},0}+\Delta \Vec{H}_\text{eff}$ and the concomitant precession for laser-induced demagnetization (left) and quasi-static expansion (right). While the demagnetization tilts $\Vec{H}_\text{eff}$ out-of-plane by adding a $\Delta \Vec{H}_\text{eff}$ in the positive $z$ direction, quasi static expansion has the opposite effect. Therefore, the induced magnetization precession starts in opposite directions for the individual mechanisms. The corresponding phase shift of $\pi$ results in a destructive superposition of the two effects. The total effective field change $\Vec{H}_{\text{eff},0}$ depends on the temperature increase $\Delta T$ and the temperature before excitation $T_\text{init}$. The quasi-static strain $\eta_\text{qs}$ depends linearly on $\Delta T$ and is insensitive to $T_\text{init}$. The demagnetization, in contrast, is non-linearly related to both $\Delta T$ and $T_\text{init}$ especially in the vicinity of $T_\text{C}$.

Figure~\ref{fig:fig_2_sim}(b) illustrates how the precession amplitude $A=\Delta \Vec{H}_\text{eff} \times \Vec{M}$ depends on the "initial" temperature $T_\text{init}$, set by a first pulse\cite{repp2020, matt2021}, and the temperature change $\Delta T$ set by a second pulse. The blueish color denotes an expansion-dominated driving of the precession (blue line in Fig.~\ref{fig:fig_2_sim}(a)) and reddish color denotes a demagnetization-dominated driving of the precession starting with an out-of-plane tilting of the magnetization (red line in Fig.~\ref{fig:fig_2_sim}(a)). The magneto-elastic coupling parameter $b_1=-8.24\cdot 10^6\,\text{J}\,\text{m}^{-3}$ has been calibrated for a similar Ni film\cite{jare2023} and the quasi-static expansion $\eta_\text{qs}=\alpha\Delta T$ is given by the ultrafast expansion coefficient $\alpha=2.02\cdot 10^{-5}\,\text{K}^{-1}$ of Ni\cite{nix1941, matt2023}. The demagnetization $\Delta M(t)$ is approximated by the temperature-dependent magnetization in thermal equilibrium $M(T)$ given by the steady state of a microscopic 3-Temperature model (m3TM) \cite{koop2010} with $T_\text{C}=630\,\text{K}$: $\Delta M(t)=M(T_\text{init}+\Delta T)-M(T_\text{init})$. Additionally, we use the saturation magnetization $M_\text{sat}=4.0\cdot 10^5 \, \text{A}\,\text{m}^{-1}$ determined by vibrating sample magnetometry.

The grey solid line in Fig.~\ref{fig:fig_2_sim}(c) represents the modelled precession amplitude $A$ for increasing $\Delta T$ at $T_\text{init}=300\,\text{K}$ that describes the fluence series in Fig.~\ref{fig:fig_1_overview}(b). Assuming $0.7\,\text{mJ}\,\text{cm}^{-2}$ to increase the phonon temperature by $\Delta T_\text{Ni}= 34.5\,\text{K}$ yields excellent agreement of our model with the fluence-dependent amplitude (colored symbols) extracted from a fit of the transient magnetization by an exponential decaying background and a damped oscillation. With increasing fluence the non-linear increase of the demagnetization reduces the increase of the precession amplitude up to $\Delta T_\text{Ni}=140\,\text{K}$ where the additional demagnetization overcomes the additional quasi-static expansion and the precession amplitude starts to decrease again. Beyond $\Delta T_\text{Ni}=270\,\text{K}$ the demagnetization even dominates over the quasi-static expansion. This path along increasing $\Delta T$ is indicated by the arrow in Fig.~\ref{fig:fig_2_sim}(b). The transition through the white area indicates the new dominant mechanism of demagnetization which can be seen directly in Fig.~\ref{fig:fig_1_overview}(b) as the $\pi$ phase shift of the transients with fluence $F=0.7$ and $5.8\,\text{mJ}\,\text{cm}^{-2}$, although the remagnetization dominating the dynamics for high fluences superimposes the oscillations.

\begin{figure*}[t!]
\centering
\includegraphics[width =0.9\textwidth]{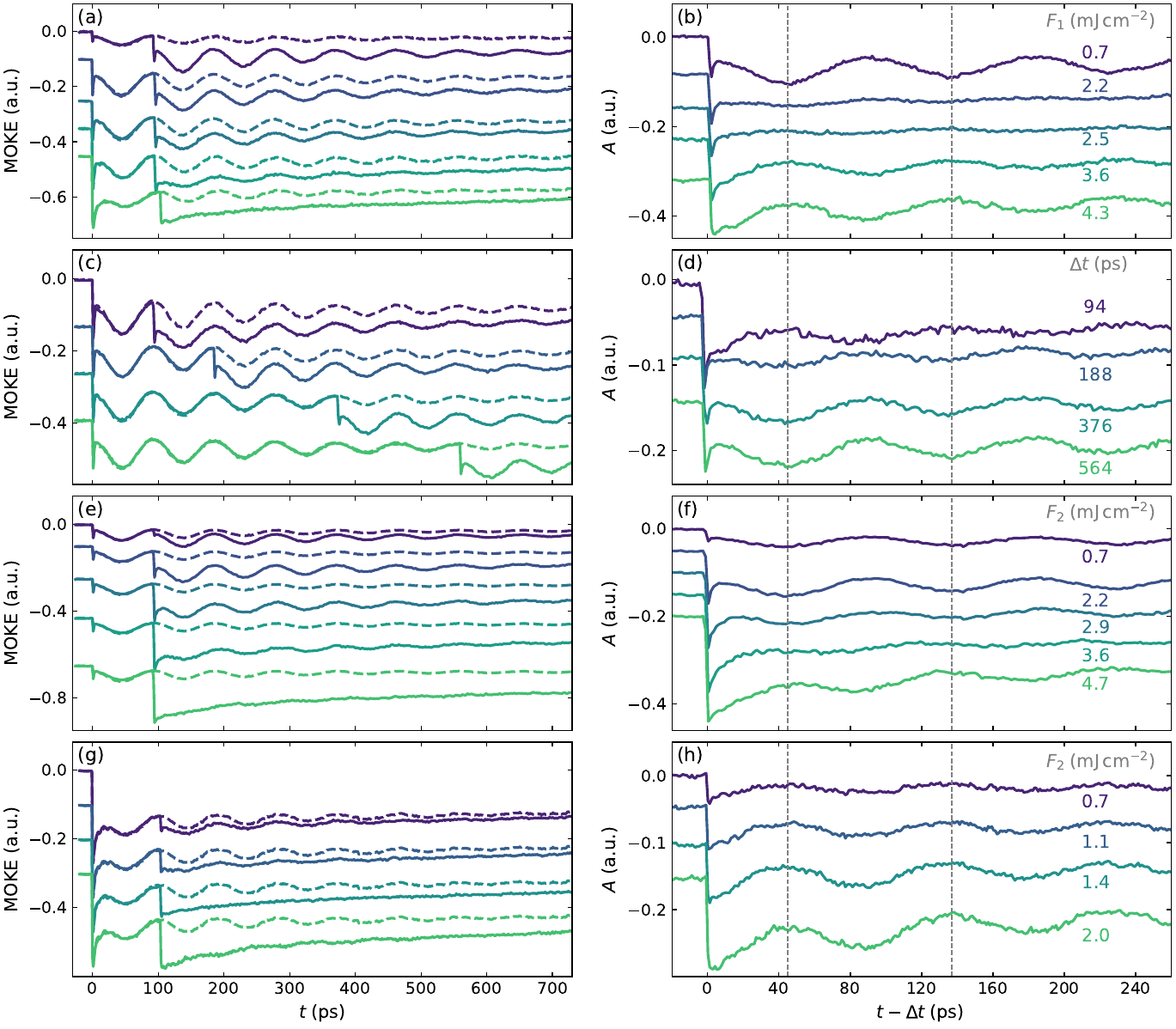}
\caption{\label{fig:fig_3_data}\textbf{Control of the magnetization precession within double-pulse excitation by tuning the effect of the second pulse:} (b,d,f,h) Difference in the transient magnetization upon both excitations (solid lines) and only the first excitation (dashed lines) in panels (a,c,e,f), respectively. (a,b) Variation of fluence of the first pulse for constant $F_2=1.8\,\text{mJ}\,\text{cm}^{-2}$. (c,d) Variation of the pump-pump delay for constant $F_1=2.8\,\text{mJ}\,\text{cm}^{-2}$ and $F_2=1.8\,\text{mJ}\,\text{cm}^{-2}$. Note that the second pulse is chosen to pump at the same precession phase at all chosen $\Delta t$ and nonetheless has an effect on the phase for varying $\Delta t$. (e,f) Variation of the fluence $F_2$ for constant $F_1=1.1\,\text{mJ}\,\text{cm}^{-2}$. (g,h) Variation of the fluence $F_2$ for constant $F_1=4.3\,\text{mJ}\,\text{cm}^{-2}$. The difference in panels (b,d,f,h) demonstrate a tuning of the dominant driving mechanism for the second excitation, which we utilize to selective control the magnetization precession upon double-pulse excitation in panels (a,c,e,f).}
\end{figure*}

Figure~\ref{fig:fig_2_sim}(b) shows that this fluence dependence of the precession amplitude strongly depends on $T_\text{init}$. With increasing $T_\text{init}$, quasi-static expansion and demagnetization compensate each other ($A=0$) at smaller $\Delta T$ and finally demagnetization always dominates for $T_\text{init}>450\,\text{K}$. We verify our model by double-pulse experiments. The temperature rise of the first pulse $\Delta T_1 \propto F_1$ raises the temperature and thus sets the initial temperature $T_{\text{init},2}=300\,\text{K}+\Delta T_1$ for the second pulse. Figure~\ref{fig:fig_3_data} presents three different double-pulse control scenarios: The control parameters for the precession amplitude are i) the fluence of the first pulse $F_1$ (Fig.~\ref{fig:fig_3_data}(a,b)), ii) the time delay $\Delta t$ between the pulses (Fig.~\ref{fig:fig_3_data}(c,d)) and iii) the fluence $F_2$ of the second pulse. For $F_2$ we show two alternatives according to the fixed pump fluence $F_1$ which either sets the initial temperature $T_\text{init,2}$ above or below the temperature $T_\text{comp}$, where the magneto-acoustic and demagnetization mechanisms compensate each other as drivers for the precession. 

The left column of Fig.~\ref{fig:fig_3_data} compares the MOKE response of the sample to the both pulses (solid lines) to the response when only the first pulse is active (dashed lines). The right column displays the difference of these two signals. The oscillation amplitude $A$ of this difference signal represents the change that is induced by the second pulse, i.e.\ the value encoded in Fig.~\ref{fig:fig_2_sim}(b). The variation of $F_1$ for constant $F_2=1.8\,\text{mJ}\,\text{cm}^{-2}$ in panels (a) and (b) corresponds to a variation of $T_{\text{init},2}$ at constant time delay and laser-induced temperature rise $\Delta T_2 =85\,\text{K}$. In agreement with our model, we observe a sign change of the oscillation amplitude $A$ in panel (b) when $T_\text{init,2}= T_\text{comp}$ for $F_1 \approx 2.4\,\text{mJ}\,\text{cm}^{-2}$. At this fluence $F_1$ the dominant driving mechanism for the second excitation switches to demagnetization. Therefore, the effect of the second pulse switches from amplifying to attenuating the precession, i.e. the amplitudes of the solid and dashed lines in Fig.~\ref{fig:fig_3_data}(a) are nearly equal. Interestingly, the rather weak second pulse ($F_2=1.8\,\text{mJ}\,\text{cm}^{-2}$) is able to switch off the precession induced by a strong first pulse with $F_1=4.3\,\text{mJ}\,\text{cm}^{-2}$ (see Fig.~\ref{fig:fig_3_data}(a)).

This selection of the dominant driving mechanism via the temperature before the second excitation $T_{\text{init},2}$ can also be realized by the pump-pump delay $\Delta t$ (see Fig.~\ref{fig:fig_3_data}(c,d)) utilizing the cooling of Ni via heat transport into the substrate within hundreds of picoseconds. With increasing $\Delta t$ for fixed $F_1=2.8\,\text{mJ}\,\text{cm}^{-2}$ and $F_2=1.8\,\text{mJ}\,\text{cm}^{-2}$, we observe the transition from demagnetization to magnetoacoustics as the driver of the precession between $\Delta t = 94$ and $188\,\text{ps}$ (Fig.~\ref{fig:fig_3_data}(d)). This is in agreement with our model, where $T_{\text{init},2}$ decreases with $\Delta t$ starting from its maximum value of $430\,\text{K}$. We chose $\Delta t$ such that the second pulse always arrives at a maximum of the precession driven by the first pulse. Thus the time delay does not have the same role as in a conventional coherent control scenario. Although the second pulse always interacts with the sample at the same phase value of the precession, the change of the dominant driving mechanism of the precession switches the effect of the second pulse from attenuation to amplification. A substrate with a phononic heat conductivity larger than glass ($\kappa_\text{glass}\approx 1\,\text{W}\,\text{m}^{-1}\,\text{K}^{-1}$), e.g. MgO, would enhance the range of attenuation and amplification that can be selected by the pump-pump delay.

Figures.~\ref{fig:fig_3_data}(e,f) demonstrate a similar tuning of the dominant driving mechanism by the fluence of the second pulse for a fixed $F_1=1.1\,\text{mJ}\,\text{cm}^{-2}$. The amplitude $A$ driven by the second pulse increases with increasing $F_2$ up to its maximum value for $F_2=2.2\,\text{mJ}\,\text{cm}^{-2}$. A further increase of $F_2$ reduces the amplitude again  and even inverts the sign. For the highest fluence, $F_2=4.7\,\text{mJ}\,\text{cm}^{-2}$, the demagnetization-dominated mechanism for the second pulse fully switches off the precession induced by the first pulse, which is best seen in Fig.~\ref{fig:fig_3_data}(e).

This dependence of the precession amplitude $A$ on the fluence $F_2$ drastically depends on the fluence of the first pulse $F_1$. Figure~\ref{fig:fig_3_data}(g,h) shows that for $F_1=4.3\,\text{mJ}\,\text{cm}^{-2}$, any choice of the fluence $F_2$ drives the precession dominantly by demagnetization, and hence reduces the precession amplitude when the second pulse excites one precession period after the first pulse that drove the precession via magneto-acoustics. This is again in full in agreement with the model (see arrow in Fig.~\ref{fig:fig_2_sim}(b)).

In summary, we demonstrated that a transition of the dominant driving mechanism of a second excitation can be achieved by the fluences of both pulses and the pump-pump delay in a double-pulse excitation scheme. Our model and experiment confirm that the demagnetization and quasi-static expansion counteract each other as precession drivers if the magneto-elastic coupling is negative ($b_1<0$). The change of the dominant driving mechanism from magneto-elastic coupling to demagnetization occurs at a fluence-dependent compensation temperature $T_\text{comp}$ below the Curie temperature $T_\text{C}$. At this point the effect of the second pulse on the precession switches from amplification to attenuation. We demonstrate that this mechanism can be exploited in various ways thanks to the non-linear fluence-dependent demagnetization mechanism, enabling coherent control beyond the standard superposition of waves.

We acknowledge the DFG for financial support via No.\ BA 2281/11-1 and Project-No.\ 328545488 – TRR 227 (projects A2 and A10). We thank Alexander von Reppert from the University of Potsdam for stimulating discussions and Juan-Carlos Rojas-S\'{a}nchez from Institute Jean Lamour in Nancy, France for determining the saturation magnetization. 

\bibliography{references.bib}
\end{document}